\documentclass[preprint2]{aastex}

\def\be{\begin{equation}}
\def\ee{\end{equation}}

\begin{document}

\title{Pulsar Efficiency} 

\author{Andrei Gruzinov}

\affil{CCPP, Physics Department, New York University, 4 Washington Place, New York, NY 10003}

\begin{abstract}

Pulsar efficiency, defined as the ratio of the pulsed bolometric luminosity to the spin-down power, is calculated to be $\approx15\%$ (averaged over the spin-dipole inclination angle, ranging between $\approx50\%$ for the aligned and $\approx10\%$ for the orthogonal). We also estimate the characteristic photon energy and argue that our results agree with the Fermi pulsar catalog -- in a sense.

~~

\end{abstract}

\section{Introduction} 

Pulsars create magnetospheres around themselves (Goldreich and Julian 1969, Ruderman and Sutherland 1975). Full quantitative description of a magnetosphere requires a kinetics input -- we need to know how the charges and photons are produced.  

It turns out that the main overall characteristics of the pulsar are calculable without any detailed knowledge of the kinetics. Here we calculate the pulsar spin-down power $L_{sd}$ and the pulsar efficiency $\epsilon$, defined as the ratio of the pulsed bolometric luminosity to $L_{sd}$.

Our results, valid to some 10\% accuracy, are as follows. The efficiency is 
\be
\epsilon \approx {0.5\over 1+5\sin ^2\theta },
\ee
where $\theta$ is the spin-dipole angle. The spin-down luminosity is 
\be
L_{sd}\approx {\mu ^2\Omega ^4\over c^3}(0.8+1.2\sin ^2\theta ),
\ee
where $\mu$ is the magnetic dipole moment and  $\Omega$ is the angular velocity. 

\begin{figure}
\plotone{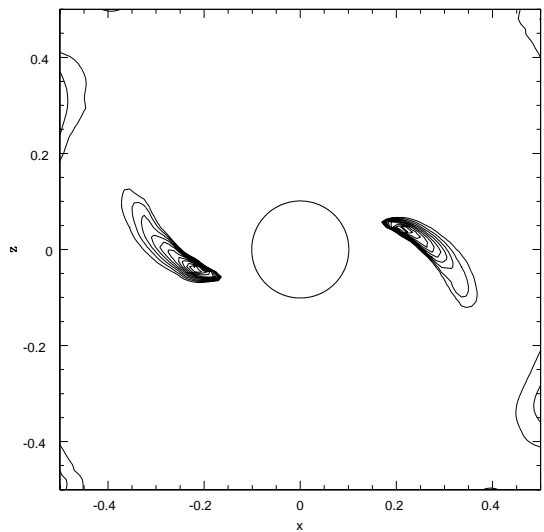}
\caption{The damping region of a pulsar with the spin-dipole angle $\theta =45^{\circ}$. Ohm's law $P=\sqrt{\rho^2+10(E^2+B^2)/r^2}$, $r$ is the spherical radius. The star rotates around the $z$ axis at half the speed of light. Shown are the isolines of $E^2-B^2$ (where positive) in the plane $y=0$. The snapshot is taken near the time when the maximum of $E^2-B^2$ passes through the plane $y=0$. The resolution is $100^3$.}
\end{figure}

\section{Observations}

(i) The Fermi pulsar catalog (Abdo et al 2010) lists 46 pulsars. Only 10 are known to have $\epsilon\lesssim 0.1$. These might be out of the beam.

(ii) Our theory (\S 3) in the present form should not be used for light curves and spectra -- kinetics seem to matter. But a crude estimate of the characteristic photon energy is possible (curvature emission in an electric field $\sim$ magnetic field at the light cylinder, radius of curvature $\sim$ light cylinder radius): $E\sim c^{3\over 8}\hbar e^{-{3\over 4}}L_{sd}^{3\over 8}\Omega ^{1\over 4}\sim 3L_{34}^{3\over 8}P_{ms}^{-{1\over 4}}{\rm GeV}$, where $L_{34}$ is the spin-down power in units of $10^{34}$erg/s and $P_{ms}$ is the pulsar period in ms. This formula is OK (to factor 3) for all the catalog pulsars but 14. For these 14 pulsars, the formula gives a frequency some 3-10 times higher than observed. All these 14 pulsars either have or may have small efficiency, and may be out of the beam.

\section{Electrodynamics of Massless Charges}

We calculate the pulsar using the so-called Electrodynamics of Massless Charges (EMC, Gruzinov 2012). EMC describes the motion of light charges in strong electromagnetic fields. Due to strong radiation damping, the dynamics becomes ``Aristotelian'', with the charge velocity rather than acceleration given by the field values. Positive and negative charges move at the speed light in the directions 
\be\label{emc}
{\bf v}_{\pm}={{\bf E}\times {\bf B}\pm(B_0{\bf B}+E_0{\bf E})\over B^2+E_0^2}.
\ee
Here the scalar $E_0$ and the pseudoscalar $B_0$ are the proper electric and magnetic fields defined by 
\be
B_0^2-E_0^2=B^2-E^2,~ B_0E_0={\bf B}\cdot {\bf E},~ E_0\geq 0.
\ee

To calculate the electromagnetic field, we numerically solve Maxwell equations. To solve Maxwell equations we need to know the current. Let $\rho_{\pm}$ be the absolute values of the charge densities of positive and negative charges. Then the EMC equation (\ref{emc}) gives 
\be
{\bf j}={\rho {\bf E}\times {\bf B}+P (B_0{\bf B}+E_0{\bf E})\over B^2+E_0^2},
\ee
where $\rho=\rho_+-\rho_-$ and $P=\rho_++\rho_-$. Since $\rho=\nabla\cdot {\bf E}$, we almost have an Ohm's law (an expression for ${\bf j}$ in terms of ${\bf E}$ and ${\bf B}$):
\be\label{ohm}
{\bf j}={(\nabla \cdot {\bf E}) {\bf E}\times {\bf B}+P (B_0{\bf B}+E_0{\bf E})\over B^2+E_0^2}.
\ee
Equation (\ref{ohm}) is not quite an Ohm's law, because $P$ is unknown (we only know that $P\geq |\rho |$). 

We have checked that when the neutral plasma density $P-|\rho |$ becomes large enough, in practice for $P-|\rho |\gtrsim |\partial B|$, the simulation results become insensitive to the actual value of $P$. This happens in the following way. In most of the volume, the ${\bf E}$ and ${\bf B}$ fields are force-free, $\rho {\bf E}+{\bf j}\times {\bf B}=0$. There is also a small volume where the fields are not force-free. In this volume some of the Poynting flux emitted by the star gets dissipated. For larger $P$, the non-force-free volume shrinks, but the total dissipated power remains fixed. The damping region is shown in Fig. 1.

\section{Appendix: Numerics }

(i) We use uniform cubic grid in a cube. ${\bf E}$ components are at the corresponding (parallel) edges of the grid, ${\bf B}$ components are at the corresponding (orthogonal) faces of the grid. We use outgoing boundary conditions at the faces of the cube. We solve the evolutionary problem 
\be\label{max} 
\partial _t {\bf B}=-\nabla \times {\bf E},~\partial _t {\bf E}= \nabla \times {\bf B}-{\bf j}.
\ee
(ii) We have a rotating star at the center of the cube. This means that inside some sphere the current ${\bf j}$ is given by the Ohm's law in a rotating frame plus an external rotating current density. The external current density is axisymmetric, with the symmetry axis of the current inclined with respect to the rotation axis. We get the dipole radiation.

(iii) Then we add the Ohm's law (\ref{ohm}) outside the star. At this stage we have to regularize equations (\ref{max}) by adding small diffusivities ($\eta \Delta {\bf B}$ and $\eta \Delta {\bf E}$ with small $\eta$). The Poynting flux emanating from the star is the spin-down power $L_{sd}$. The Poynting flux leaving the simulation cube is $(1-\epsilon)L_{sd}$.

\end{document}